\documentclass[a4paper]{jpconf}
\usepackage{graphicx}
\begin{document}
\title{A Generalized Spin Statistics Theorem}

%\vspace{5mm} Submitted: March 30, 2012}
%\end{center}
\author{Paul O'Hara}
\address{Istituto Universitario Sophia,Via San Vito, 28 - Loppiano, 50064 Figline e Incisa Valdarno (FI), Italy}
\ead{paul.ohara@iu-sophia.org}
\newtheorem{thm}{Theorem}
\newtheorem{cor}{Corollary}
\newtheorem{Def}{Definition}
\newtheorem{lem}{Lemma}
\begin{abstract}
In this article we generalize the spin statistics theorem and show that a state obeys Fermi-Dirac statistics if and only if
the state is invariant under the action of $SL(n,\mathcal{C})$. We also briefly discuss the experimental evidence and how the theorem relates to spin entanglement. 
\end{abstract}

\section{Introduction}
The origin of quantum statistics seems to have begun in 1920 when S.K. Bose sent a paper to Einstein seeking his help in getting it published. Einstein recommended it to Zeitschrift but later also published his own version in which the notion of indistinguishable photon states were introduced \cite {streat}.
This was the beginning of what is now referred to as Bose-Einstein statistics.  Another development took place in 1925 with the formulation of the Pauli exclusion principle which asserts that no two electrons in an atom could be in the same quantum state.  In the 1930's this was subsequently generalized by Fermi and Dirac into what is now referred to as Fermi-Dirac statistics \cite{brit}. At about the same time Jordan and Wigner second quantized the Schrodinger equation and showed that Bose-Einstein statistics and Fermi-Dirac statistics respectively obeyed a set of commutator relations and anti-commutator relations applied to creation and annihilation operators \cite{streat}.  This was the precursor of a connection between spin and statistics, first formulated by Markus Fiertz in 1939 \cite{fierz}, and then further developed by Pauli a year later \cite{pauli}. In his paper Pauli claims \textit{the necessity of Fermi-Dirac statistics for particles with arbitrary half-integral spin}, and of \textit{the necessity of Einstein-Bose statistics for particles with arbitrary integral spin}. Also by invoking relativistic invariance he shows that bosons cannot be quantized as fermions and vice-versa.

Most subsequent work on spin-statistics takes for granted Pauli's conclusions but also struggles to understand the physical (as opposed to the mathematical) principles involved.  For example, Feynman in his Lecture in Physics series states: ...\textit{An explanation has been worked out by Pauli from complicated arguments of QFT and relativity...but we haven't found a way of reproducing his arguments on an elementary level...this probably means that we do not have a complete understanding of the fundamental principle involved...}\cite{Feyn} Indeed within the context of Feynman's obsrvation, Duck and Sudarshan give a comprehensive analysis of the many different approaches to spin-statistics including work by De Wet, Wightman, Schwinger, Feynman, Hall, Luder and Zumino and conclude that the various proofs, including their own, \textit{[were] not completely free from the complications of relativistic quantum field theory} \cite{DS}. Also Berry and Robbins article on the subject published in 1997 cannot be considered \textit{elementary} in Feynman's sense \cite{romer}, \cite{BR}.    With this in mind, in Theorem 1  we prove another version of the spin-statistics theorem which is free of quantum field theory complications.  We show that Fermi-Dirac statistics is directly related to $SL(n,\cal{C})$  invariance.

However, before doing so, we note that this result is itself a generalization of  Theorem 2 in \cite{ohara} where it was previously shown that the rotational invariance associated with the existence of pairwise entangled states was  sufficient for the Pauli exclusion principle. It is also suggested both in \cite{ohara} and \cite{ohara2} that pairwise entanglement can be used to explain the stability of spin-$\frac{1}{2}$ baryons. In other words, `` spin-$\frac{3}{2}$ baryons may be viewed as excited states of spin-$\frac{ 1}{2}$ baryons"\cite{ohara} which will decay into a stable spin-$\frac{1}{2}$ proton. 

\section{A spin statistics theorem}  
The importance of this paper is not the discussion about entanglement per se but rather the proof of Theorem 1 which states that a necessary and sufficient condition to have Fermi-Dirac statistics is invariance under the action of  $SL(n,\cal{C})$. The rotational invariance is embedded in the observation  that $SU(2,{\cal{C}})\subset SL(n,\cal{C})$. The theorem is very general. It applies to any tensor product vector space of the form $V=V_1\otimes \cdots \otimes V_n$. In particular if we choose a vector of the form
$$v\equiv v_1\wedge v_2\wedge \dots \wedge v_n,\  \  v_i= (v_{ij}), 1\le j \le n$$  where the wedge product indicates an anti-symmetric vector then

\begin{eqnarray*} v&=&\left(\begin{array}{c}v_{11}\\\vdots\\v_{n1} \end{array}\right) \wedge \left(\begin{array}{c}v_{12}\\\vdots\\v_{n2} \end{array}\right)\wedge
\dots \wedge \left(\begin{array}{c}v_{n1}\\\vdots \\v_{nn} \end{array}\right)\\
&=&\left|
     \begin{array}{cccc}
       v_{11} & v_{12} & \cdots & v_{1n} \\
       v_{21} & v_{22} & \cdots & v_{2n} \\
       \vdots & \vdots & \vdots & \vdots \\
       v_{n1} & v_{n2} & \cdots & v_{nn} \\
     \end{array}
   \right|{\bf e}_1\wedge {\bf e}_2\wedge \dots \wedge{\bf e}_n\\
&=& |v|{\bf e}_1\wedge {\bf e}_2\wedge \dots \wedge{\bf e}_n,\ {\rm where}\   |v|=\left|
     \begin{array}{cccc}
       v_{11} & v_{12} & \cdots & v_{1n} \\
       v_{21} & v_{22} & \cdots & v_{2n} \\
       \vdots & \vdots & \vdots & \vdots \\
       v_{n1} & v_{n2} & \cdots & v_{nn} \\
     \end{array}
   \right|
\end{eqnarray*}
$|v|$ is usually called the Slater determinant and it remains invariant for any choice of orthonormal basis $\{{\bf e}_1, {\bf e}_2 \dots {\bf e}_n \}$  of $V_i$. Historically, in the physics literature, the Slater determinant has always been associated with Fermi-Dirac statistics, and used to characterize the anti-symmetric nature of the wavefunction. 

As a consequence of the invariance of the Slater determinant defined with respect to an orthonormal basis, it follows that if we take any matrix  element, $T$ of the group $SL(n, {\cal C})$ (which by definition is the group of all elements with determinant 1), and apply this operator to each component of the antisymmetric vector $v$ then  
$$Tv=|Tv|{\bf e}_1\wedge {\bf e}_2\wedge \dots \wedge{\bf e}_n=|T|v||{\bf e}_1\wedge {\bf e}_2\wedge \dots \wedge{\bf e}_n=v .$$   Moreover, as the second theorem notes, the antisymmetric tensor $v$ is the only vector with this property. Consequently, the two theorems taken together suggest that Fermi-Dirac statistics for n indistingushable particles be formally  defined as any statistic that is invariant under the action of $SL(n, {\cal C})$. The formal proofs are presented below.   
  
\begin{thm} Let $V=V_1\otimes \cdots \otimes V_n$, where for all $i$, $j$, each $V_i\cong V_j$ and $V_i$ is an n-dimensional vector space.
Let $T=T_1\otimes \cdots \otimes T_n$
where for each $i,j,\ T_i = T_j$ and $T_i$ is a linear operator on $V_i$. Let
\begin{eqnarray*}v&\equiv& v_1\wedge v_2\wedge \dots \wedge v_n\\
&=&\left(\begin{array}{c}v_{11}\\\vdots\\v_{n1} \end{array}\right)\wedge \left(\begin{array}{c}v_{12}\\\vdots\\v_{n2} \end{array}\right)\wedge
\dots \wedge \left(\begin{array}{c}v_{1n}\\\vdots\\v_{nn}\end{array}\right)\end{eqnarray*}
then for $v\ne 0$
$$Tv=v\ \iff \ T\in \bigotimes^n_1{SL(n,\mathcal{C})}.$$
In other words, Fermi-Dirac statistics is invariant under the action of $SL(n,\mathcal{C})$.
Note by definition $v_1\wedge v_2\wedge \dots \wedge v_n = \frac{1}{n!}\delta^{i_1\dots i_n}_{1\dots n}v_{i_1}\otimes \dots \otimes v_{i_n}$
\end{thm}
{Proof:} Let $\{{\bf e}_1, {\bf e}_2 \dots {\bf e}_n \}$ be an orthonormal basis of $V_i$, then
\begin{eqnarray*} v&=&\left(\begin{array}{c}v_{11}\\\vdots\\v_{n1} \end{array}\right) \wedge \left(\begin{array}{c}v_{12}\\\vdots\\v_{n2} \end{array}\right)\wedge
\dots \wedge \left(\begin{array}{c}v_{n1}\\\vdots \\v_{nn} \end{array}\right)\\
&=&\left|
     \begin{array}{cccc}
       v_{11} & v_{12} & \cdots & v_{1n} \\
       v_{21} & v_{22} & \cdots & v_{2n} \\
       \vdots & \vdots & \vdots & \vdots \\
       v_{n1} & v_{n2} & \cdots & v_{nn} \\
     \end{array}
   \right|{\bf e}_1\wedge {\bf e}_2\wedge \dots \wedge{\bf e}_n\\
&=& |v|{\bf e}_1\wedge {\bf e}_2\wedge \dots \wedge{\bf e}_n,\ {\rm where}\   |v|=\left|
     \begin{array}{cccc}
       v_{11} & v_{12} & \cdots & v_{1n} \\
       v_{21} & v_{22} & \cdots & v_{2n} \\
       \vdots & \vdots & \vdots & \vdots \\
       v_{n1} & v_{n2} & \cdots & v_{nn} \\
     \end{array}
   \right|
\end{eqnarray*}
The linearity of $T$ gives
\begin{eqnarray*} Tv&=& |v|T_1{\bf e}_1\wedge T_2{\bf e}_2\wedge \dots \wedge T_n{\bf e}_n\\
&=&|v|\left(\begin{array}{c}t_{11}\\\vdots\\t_{n1} \end{array}\right) \wedge \left(\begin{array}{c}t_{12}\\\vdots\\t_{n2} \end{array}\right)\wedge
\dots \wedge \left(\begin{array}{c}t_{n1}\\\vdots \\t_{nn} \end{array}\right)\\
        &=&|v||T_1|{\bf e}_1\wedge {\bf e}_2\wedge \dots \wedge {\bf e}_n,\qquad T_1=T_2=\dots =T_n
\end{eqnarray*}
Therefore, since $v\ne 0$ implies $|v|\ne 0$ then
$$Tv=v \Rightarrow |T_1|=1\ {\textrm{and}}\ T_1\in {SL(n,\mathcal{C})}$$
Conversely $$\ T_1\in {SL(n,\mathcal{C})} \Rightarrow Tv=v$$
This proves the theorem. 
\\

As mentioned in the introduction, this result can be seen as a generalization of a theorem where Fermi-Dirac statistics can be derived using rotational invariance \cite{ohara}. The first thing to note is that $SU(n,{\mathcal{C}}) \subset SL(n,\mathcal{C})$ and therefore the Fermi-Dirac statistic is automatically rotationally invariant.  In itself this already gives us a deeper insight into Fermi-Dirac statistics. The fact is $SU(2,{\cal C})$ and $SO(2,{\cal C})$ groups are subgroups of $SL(n,{\cal C})$ and consequently  particles which are  invariant under the action of these groups are pairwise entangled. This means that singlet states become the building blocks of Fermi-Dirac statistics.  For example, in two dimensions if we let 

$$R(\theta)= \left(\begin{array}{cc}
       \cos c\theta & \sin c\theta  \\
       -\sin c\theta & \cos c\theta \\
      \end{array}\right) \in SL(2, \cal{C})$$   
then direct calculation shows that ${R(\theta)\bf e}_1\wedge {R(\theta)\bf e}_2= {\bf e}_1\wedge {\bf e}_2 $. 

It is also important to note that $2{\bf e}_1\wedge {\bf e}_2 = {\bf e}_1\otimes {\bf e}_2 - {\bf e}_2\otimes {\bf e}_1$  represents a singlet state and is therefore entangled by definition. In the case of $n$ dimensions if
\begin{equation}
R_{ij}(\theta)=\begin{array}{c}
    \ \\
     \ \\
      \ \\
      i \\
      \ \\
      j \\
      \ \\
	\ \\
      n \\
      \end{array}
             \left(
             \begin{array}{ccccccc}
             \ &\ &i &\ &\ j\ &\ &  n \\
1 & \cdots & 0 & \cdots &0 &\cdots & 0  \\
                \vdots &\ddots & \vdots &\ddots &\vdots &\ddots & \vdots\\
                0 & \cdots & \cos(c\theta) &\cdots & \sin(c\theta) &\cdots  & 0\\
                \vdots & \ddots & \vdots &\ddots & \vdots &\ddots & \vdots\\
                0 & \cdots & -\sin (c\theta) & \cdots &\cos (c\theta) & \cdots & 0\\
                \vdots & \ddots & \vdots &\ddots & \vdots &\ddots & \vdots\\
                0 & \cdots & 0 & \cdots &0 & \cdots & 1\\
\end{array}\right)\in {SL(n,\mathcal{C})}\label{R}
\end{equation}
then
$${\bf e}_1 \wedge \dots  R_{ij}(\theta){\bf e}_i\wedge \dots \wedge R_{ij}(\theta){\bf e}_j\wedge {\bf e}_n = {\bf e}_1 \wedge \dots  {\bf e}_i\wedge \dots \wedge {\bf e}_j\wedge {\bf e}_n \ .$$ 
This captures the pairwise rotational invariance associated with the $ij$ singlet state represented by the wedge product.
 Moreover, since ${\bf e}_1\wedge {\bf e}_2 \dots \wedge {\bf e}_n $ is associative this $n$-fold state can be interpreted as being built from pairwise entangled states.\\ 

There is a second theorem closely related to the first. It is a uniqueness theorem affirming that only Fermi-Dirac states are invariant under the action of $SL(n,\mathcal{C})$.  Its proof requires the following lemma:

\begin{lem}Let $V=V_1\otimes \dots \otimes V_n$ as in Theorem 1. Let $\{e_1, \dots, e_n\}$ be an orthonormal basis of $V_i$ and $R_{ij}$ be as in equation (\ref{R}).  If $R_{ij}u=u$ where
$$u=\sum_{\sigma(1)\cdots \sigma(n)\in S_n} c_{\sigma(1)\cdots \sigma(n)}e_{\sigma(1)}\otimes \cdots \otimes e_{\sigma(n)}\ ,$$
$S_n$ is the permutation group and $c_{\sigma(1)\cdots \sigma(n)}$ are constants then $$c_{\sigma(1)\cdots i \cdots j\cdots \sigma(n)}=-c_{\sigma(1)\cdots j \cdots i\cdots \sigma(n)}$$
\end{lem}
{Proof:} If $u=0$ then the lemma follows trivially, Assume $u\ne 0$.
Note that in the case $n=2$ 
\begin{eqnarray*}&&R_{12}(c_{12} {\bf e_1}\otimes {\bf e_2} +c_{21} {\bf e_2}\otimes {\bf e_1})\\
&=& (c_{12}+c_{21})\cos \theta \sin \theta {\bf e_1 \otimes \bf e_1}+(c_{12}\cos^2 \theta  - c_{21}\sin^2 \theta){\bf e_1} \otimes {\bf e_2}\\
& & -(c_{12}\sin^2 \theta  - c_{21}\cos^2 \theta){\bf e_2} \otimes {\bf e_1}- (c_{12}+c_{21})\cos \theta \sin \theta {\bf e_2 \otimes \bf e_2}\\
&=&  c_{12} {\bf e_1}\otimes {\bf e_2} +c_{21} {\bf e_2}\otimes {\bf e_1}\qquad {\rm by\  assumption.}
\end{eqnarray*}
It follows by the linear independence of ${\bf e_1}$ and ${\bf e_2}$ that $c_{12}=-c_{21}$.

To extend this to the $n$ dimensional case, note that $R_{ij}{\bf e_k}={\bf e_k}$ in the case of $k\ne i$ and $k\ne j$.
Direct calculation gives
\begin{eqnarray*}
&\ &(R(ij)\otimes \cdots \otimes R(ij)) u\\
&=&\sum_{\sigma(1)\cdots i\cdots j\cdots \sigma(n)\in S_n} c_{\sigma(1)\dots \sigma(n)}Re_{\sigma(1)}\otimes\cdots Re_i \cdots \otimes Re_j \cdots \otimes Re_{\sigma(n)}\\
&=&\sum_{\sigma(1)\cdots i\cdots j\cdots \sigma(n)\in S_n} c_{\sigma(1)\dots \sigma(n)}e_{\sigma(1)}\otimes\cdots Re_i \cdots \otimes Re_j \cdots \otimes e_{\sigma(n)}\\
&=& u\qquad \textrm{since}\ u\ \textrm{is an eigenvector with eigenvalue 1.}
\end{eqnarray*}
Mathematically, this is equivalent to the $n=2$ case already worked out above.
It follows from linear independence that
$$ c_{\sigma(1)\cdots i\cdots j\cdots \sigma(n)}=-c_{\sigma(1)\cdots j\cdots i\cdots \sigma(n)}$$
The result has been proven. \\
\\
We now state and prove the theorem:
\begin{thm} Let $V=V_1\otimes \cdots \otimes V_n$ as in Theorem 1, and
$T=T_1\otimes \cdots \otimes T_n$
where for each $i,j,\ T_i=T_j$ and $T_i(\theta)\in SL(n,\mathcal{C})$. If for all $T\in \bigotimes^n SL(n,\mathcal{C})$,
$Tv=v,\ v\ne 0$  then
\begin{equation}
v= \kappa(v_1\wedge v_2\wedge \dots \wedge v_n), \ \kappa {\rm \ an\ arbitrary\ constant.}\label{thm}
\end{equation}
\noindent This means that if $v\ne 0$ is invariant under the action of $SL(n,\mathcal{C})$ and $\kappa =1$ then it must be a Fermi-Dirac statistic.
\end{thm}
{Proof:} We need to show that if $v$ is invariant under the action of any operator $T\in \bigotimes^n SL(n,\mathcal{C})$ then
$v$ is given as in (\ref{thm}).
Indeed, from Theorem 1, we know that the Fermi-Dirac state is invariant under the action of $SL(n,\mathcal{C})$. It remains to show that it is unique upto a multiplicative constant.\\
\\
In general if $u\in V\otimes \cdots \otimes V$ then
\begin{equation}u=\sum_{i_1\dots i_n\in \aleph_n} c_{i_1\dots i_n}e_{i_1}\otimes \cdots \otimes {e}_{i_n}\label{lincomb1}\end{equation}
where the $\aleph_n =\{1, \cdots , n\}$ and $\{{e}_{i_1}\otimes \cdots \otimes {e}_{i_n}|i_j\in \aleph_n\}$ forms a basis for the space. Note that there are $n^n$ summed terms in equation (\ref{lincomb1}).
It remains to show that if $Tu=u$ for an arbitrary $T$ then $u=v$.  This is achieved by showing that the action of suitably chosen elements of
$SL(n,\mathcal{C})$ on $u$ impose restrictions on equation (\ref{lincomb1}) until only $v$ remains.\newline
\\
In particular, for the Lie group $\{\exp(\theta J)|tr(J)=0\}\in SL(n,{\mathcal{C}})$, if we let
$L(\theta)=(L_1(\theta)\otimes \dots \otimes L_n(\theta))$, where each $L_i(\theta)=\exp(\theta J)$
there exists a complete set of eigenvectors $\{{\bf e}_1, {\bf e}_2 \dots {\bf e}_n \}$ of $L_i$, forming a basis for $V_i$, with eigenvalues $e^{\lambda_1\theta}, e^{\lambda_2 \theta}, \dots e^{\lambda_n\theta}$ such that \cite{FS}
\begin{equation}\lambda_1 +\lambda_2+\dots +\lambda_n=0 \label{roots}\end{equation}
and  $L_i=diag\{{\bf e}_1, {\bf e}_2 \dots {\bf e}_n \}.$
It is clear that for every permutation $\sigma \in S_n$, where $S_n$ is the permutation group, the set of tensor products
$$\{{\bf e}_{\sigma(1)}\otimes {\bf e}_{\sigma(2)}\otimes \dots \otimes {\bf e}_{\sigma(n)}\}$$
characterize a basis for all independent eigenvectors of $L(\theta)=L_1\otimes \dots \otimes L_n$ with eigenvalue 1. 
Indeed, all other linearly independent eigenvectors of $L(\theta)$ can be expressed in the form 
$${\bf e}_{\sigma(1)}\dots \otimes {\bf e}_{i}\otimes \dots {\bf e}_{i}\dots  \otimes {\bf e}_{\sigma(n)}, \qquad {\bf e}_i \ne {\bf e}_j$$
with 
$$\lambda_1+ \dots \lambda_i\dots \lambda_i\dots +\lambda_n \ne \lambda_1+ \dots \lambda_i\dots \lambda_j\dots +\lambda_n =0$$
unless $\lambda_i=\lambda_j$. However we have chosen $L$ such that each $\lambda_i$ is a distinct $n$-th root of unity and therefore $\lambda_i\ne \lambda_j$.
It follows that equation (\ref{lincomb1}) under the action of $L(\theta)$ reduces to
\begin{equation} u=\sum_{\sigma(1)\dots \sigma(n)\in S_n} c_{\sigma(1)\dots \sigma(n)}e_{\sigma(1)}\otimes \cdots \otimes e_{\sigma(n)}\label{lincomb2}\end{equation}
To conclude the proof, we turn to the lemma. Let $T=R_{ij}$ be as above. Note $R(ij)\in SL(n,{\mathcal{C}})$. Invoking  the lemma now requires that
$$c_{\sigma(1)\dots \sigma(i)\dots \sigma(j)\dots \sigma(n)} = -c_{\sigma(1)\dots \sigma(j)\dots \sigma(i)\dots \sigma(n)}$$
for every $i\neq j$. This  gives $u=v$.
\noindent The theorem has been proven. \\
\\
The above theorem applies to any n-dimensional vector space with an n-fold tensor product defined on it. We now extend this to include an n-fold vector space with an m-fold tensor product ($n\ge m$)
\begin{cor}Let $V=V_1\otimes \cdots \otimes V_m$, where each $V_i$ is an n-dimensional vector space ($m\le n$), and $W_i\subset V_i$ an $m$-dimensional subspace.
Let $T=T_1\otimes \cdots \otimes T_m$
where for each $i,j,\ T_i=T_j$ and $T_i$ is a linear operator on $V_i$ leaving $W_i$ invariant, i.e. $T_i=T_{iW}\oplus T_{i(V-W)}$, with the understanding that $T_{iW}$ is the operator $T$ restricted to the subspace $W$. If
\begin{eqnarray*}v\equiv v_1\wedge v_2\wedge \dots \wedge v_m \in W\ {\rm and}\ v\ne 0
\end{eqnarray*}
then
$$T_{W}v=v\ \iff \ T_{W}\in \bigotimes^m_1{SL(n,\mathcal{C})}.$$
In other words, Fermi-Dirac statistics restricted to a subspace is invariant under the action of $SL(n,\mathcal{C})$ restricted to the same subspace.
\end{cor}
{Proof:} $T=T_{W}\oplus T_{V-W}$ The proof then follows by applying Theorem 1 to $T_W$ and noting that $T_W$ is restricted to $W$.

\section{Bose-Einstein statistics}
Based on the above, an alternative definition of a Fermi-Dirac statistics can be given:
\begin{Def} In a tensor product space of the form $V_1\otimes \dots \otimes V_n$, where each $V_i\cong V_j$, a Fermi-Dirac statistic is a state which is invariant under the action of the group $SL(n,{\cal C})$.
\end{Def}
Theorem 2 affirms that once a normalization is chosen such a state is unique. Moreover, in order to generate non Fermi-Dirac statistics, it is sufficient to relax the conditions specified by the definition.  Specifically, in keeping with the usual definition we can define Bose-Einstein statistics as follows:
\begin{Def} In a tensor product space of the form $V_1\otimes \dots \otimes V_n$, where each $V_i\cong V_j$, a Bose-Einstein statistic is a state which is invariant under the action of the permutation group $S_n$.
\end{Def}
It is important to note that both the Fermi-Dirac and Bose-Einstein states are invariant under the action of the set of even permutations $A_n\subset S_n$.  However, in the case of the Fermi-Dirac statistic the invariance under the action of $A_n$ is not per se sufficient to have such states. We also require the invariance under the action of $SL(n,{\cal C})$, which as we have already previously noted is connected with the presence of spin singlet states.   This means that from the perspective of physics, Fermi-Dirac statististics can be understood as the statistics of n-indistinguisable particles forming spin singlet states, while Bose-Einstein statistics can be understood as the statistics of n-indistinguisable particles where the spin singlet state dependency has been broken. In the case of Bose-Einstein statistics the spin states of indistinguisable particles are independent of each other. We  now express this observation in the following lemma and corollary. It is also worth pointing out that in the case of a statistic which is neither Fermi-Dirac or Bose-Einstein, the invariance under $A_n$ is violated. An example of this is also given below.
\begin{lem} Let $\sigma \in S_n$ be a permutaion of (1,\dots ,n), with the identity permutation denoted by id.  If
$${\bf v}=\sum_{\sigma }c_{\sigma }(v_{\sigma (1)}\dots \otimes v_{\sigma (n)})\qquad such\ that\ c_{id}=\frac{1}{n!}\  {\rm and } \  c_{\sigma} =\pm \frac{1}{n!} \ otherwise$$
is defined on the space $V_1\otimes \dots \otimes V_n$, and  is invariant under the action of $A_n$ then ${\bf v}$ obeys either
the Fermi-Dirac or the Bose-Einstein statistic.\\
\end{lem}
{\bf Proof:} Let 
\begin{eqnarray*}{\bf v}_0 &=& \sum_{\sigma\in A_n }\sigma(c_{id}(v_1
\otimes v_2\dots \otimes v_n))\\
{\bf v}_1 &=& \sum_{\sigma \in A_n}\sigma(c_{id}(v_2\otimes v_1\dots \otimes v_n))
\end{eqnarray*}
where $\sigma(c_{id}(v_1\otimes \dots \otimes v_n)\equiv c_{id}(v_{{\sigma_i}(1)}\otimes \dots \otimes v_{{\sigma_i}(n)})$.
This means that ${\bf v}_0$ and ${\bf v}_1$ are invariant by construction under the action of $A_n$, since they are respectively the sum of all even and odd permutations of $v_1\otimes \dots \otimes v_n$ . Therefore, the invariance of ${\bf v}$ with respect  to $A_n$ requires that 
$${\bf v} - ({\bf v}_0 \pm {\bf v}_1)$$
is also invariant. By using linear independence we find that this can only occur if 
$${\bf v} = ({\bf v}_0 +{\bf v}_1)\qquad {\rm or}\qquad {\bf v} = ({\bf v}_0 -{\bf v}_1)$$
which define the Bose-Einstein and Fermi-Dirac statistics respectively. The result follows.  
%\vskip 7pt

\begin{cor} Let ${\bf v}$ be as above
such that no two particles are in a singlet state then ${\bf v} = ({\bf v}_0 + {\bf v}_1)$, which means
this  system of particles obeys the Bose-Einstein statistics.
\end{cor}
{\bf Proof:} Since ${\bf v}$ is invaraint under the action of $A_n$ then it must be either a Fermi-Dirac or Bose-Einstein statistic. However, there are no singlets, and so it cannot be invariant under $SL(2, C)\subset SL(n,C)$. Therefore, it cannot  be a Fermi-Dirac statistic by definition. Therefore, it obeys Bose-Einstein statistics. 
\vskip 7pt
Inherent in this lemma and its corollary is the fact that Fermi-Dirac statistics requires not only indistinguishability but also
that the particles form singlet states. In other words, Fermi-Dirac statistics presuposses that particles are entangled and consequently dependent on each other while Bose-Einstein statistics is a consequence of breaking the entanglement. Within the context of atoms or molecules this entanglement can be associated with the electron orbitals. 

It might be instructive to apply the above theorem to a three particle wave
function that is {\it not} of the above type. Consider:
\begin{eqnarray*} {\bf v}&=&v_1\otimes (v_2\otimes v_3+v_3\otimes v_2)+v_2\otimes (v_3\otimes v_1+v_1\otimes v_3)\\
&\  & \qquad +v_3\otimes (v_1\otimes v_2- v_2\otimes v_1)
\end{eqnarray*}
On putting $v_1=v_2$,
\begin{eqnarray*}{\bf v}
= v_1\otimes
(v_2 \otimes v_3+v_3 \otimes v_2)+v_2 \otimes (v_3\otimes v_1+v_1\otimes v_3)
\end{eqnarray*}
which is not invariant under $A_3$ and a fortiori $SL(3,{\cal C})$. It is also not invariant under $S_3$. 

The above theorems and lemma also implicitely explain how to construct various types of parastatistics.
For example the five electrons in the boron atom obey the Fermi-Dirac statistics associated with $SL(5, {\cal C})$ invariance. On the other hand, if we consider the two electrons of the helium atom together with the three electrons of the lithium atom then theses electrons obey $SL(2, {\cal C})\otimes SL(3,{\cal C})$ statistics. The process of assigning electrons to different atoms partially distinguishes them. 

\section{Relationship to Special Relativity and QFT}   
The above theorems and corollary suggest a general criteria for classifying Fermi-Dirac statistics and Bose-Einstein statistics. In order to complete the transition, we first need to establish some algebraic connections between the tensor formalism and matrix representations. Indeed, if we impose
some further structure on the tensor products, we can relate the above theorem to relativity and quantum field theory. 

\subsection {A Clifford Algebra Approach}
Given the relationship between the Pauli spin matrices and a Clifford Algebra,
we begin with a 2-component spinor of the group $SL(2,\mathcal{C})$ such that $\phi^{\prime}=S(l)\phi$, where
$S(l)=e^{\omega_{ab}\sigma_{ab}}\in SL(2,\mathcal{C})$, $\sigma_a=(1,\vec{\sigma})$ forms a basis for the Clifford  algebra and the vector
$x_a=\phi^{\dag} \sigma_a \phi$ is a Lorentz 4-vector of a massless particle \cite{oraif}. Moreover, any vector in this space can be expressed as $X=x^a\sigma_a$.
It should be noted that the restriction $a, b \in\{1,2,3\}$ means $\{\sigma_a,\sigma_b\}=0$ and $[\sigma_a,\sigma_b]=2i\sigma_c$. Oftentimes, physicists prefer to work with the matrices ${\bf S}_i=(\hbar/2)\sigma_i$, which are called the Pauli spin matrices. However, for this paper it is more convenient to work with $\sigma_i$, and we will call these the Pauli matrices.

In general if $S(l)\in SL(2,\mathcal{C})$ with adjoint $S^{\dagger}$ and $X$ is a hermitian (non-singular) $2 \times 2$ matrix then
\begin{equation}X^{\prime}=SXS^{\dagger} \label{boost}\end{equation}
is a transformation mapping the vector $X$ into the vector $X^{\prime}$ \cite{FS} and $det(X^{\prime})=det(X)$. We now prove the following lemma.
\begin{lem} Let $X$ and $X^*$ represent a (hermitian) 4 vector and its conjugate defined respectively by
$$X=\left(
       \begin{array}{cc}
       x_0-x_1 & x_2+ix_3 \\
        x_2-ix_3 & x_0+x_1 \\
       \end{array}
     \right),\ X^{*}=\left(
       \begin{array}{cc}
         x_0+x_1 & -x_2-ix_3 \\
         -x_2+ix_3 & x_0-x_1 \\
       \end{array}
     \right)$$
If $S(l)\in SL(2,\mathcal{C})$ and $T=(S^{\dag})^{-1}$
then $S$ and $T$ preserve conjugacy. In other words,
$$X^{\prime}= SXS^{\dagger}= \left(
       \begin{array}{cc}
         x^{\prime}_0-x^{\prime}_1 & x^{\prime}_2+ix^{\prime}_3 \\
         x^{\prime}_2-ix^{\prime}_3 & x^{\prime}_0+x^{\prime}_1 \\
       \end{array}\right)$$ implies
$$ X^{\prime*}= \left(
       \begin{array}{cc}
         x^{\prime}_0+x^{\prime}_1 & -x^{\prime}_2-ix^{\prime}_3 \\
         -x^{\prime}_2+ix^{\prime}_3 & x^{\prime}_0-x^{\prime}_1 \\
       \end{array}\right)= TX^*T^{\dagger}$$
\end{lem}
{Proof:} A simple calculation shows   $X^*=X^{-1}det(X)$.
Therefore $$X^{\prime*}=det(X^{\prime})(X^{\prime})^{-1}=det(X^{\prime})(SXS^{\dagger})^{-1}=det(X)(S^{\dag})^{-1}X^{-1}S^{-1}=TX^*T^{\dag}$$
The lemma is proven.\\
Remark:\begin{itemize}
\item It immediately follows from the lemma that
$$X^{\prime}X^{\prime*}=XX^{*}=x^2_0-x^2_1-x^2_2-x^2_3$$
       is Lorentz invariant.
\item In general $X^{\prime*}\ne X^{*\prime}$. As a counter example consider
$$X^{*\prime}=SX^{*}S^{\dag}\ \textrm{where}\  S=\left(\begin{array}{cc}
         e^{\omega} & 0 \\
         0 & e^{-\omega} \\
       \end{array}\right)$$
\end{itemize}
This means   that while $X^{*\prime}$ is a vector it is {\bf not} necessarily the conjugate vector of $X^{\prime}$. Indeed, in order for  $X^{\prime*}= X^{*\prime}$, $S$ needs to be an element of $SU(2)$ which is a subgroup of
$SL(2,\cal{C})$. We express this as a corollary.

\begin{cor} If $S\in SU(n)$ then $T=S$
\end{cor}
{Proof:} By definition of $SU(n)$, $S^{\dag}=S^{-1}$ and therefore $T=(S^{\dag})^{-1}=S$. The result follows.
\\
\subsection{Conjugate Solutions and Majorana Fields}
There is a remarkable connection between the conjugate states $X$ and $X^*$ and the solutions to the Majorana equations.  Recall that for a free particle
the Majorana equations are given by
\begin{equation} {\mathcal D}\eta=\sigma_{\mu}\partial^{\mu}\eta=-m\chi\qquad {\rm and}\qquad {\mathcal D}^*\chi=\sigma^*_{\mu}\partial^{\mu}\chi=-m\eta \label{MJ}\end{equation}
where $$\mathcal{D}=\left(
       \begin{array}{cc}
       \partial_0-\partial_1 & \partial_2+i\partial_3 \\
        \partial_2-i\partial_3 & \partial_0+\partial_1 \\
       \end{array}
     \right),\ \mathcal{D}^{*}=\left(
       \begin{array}{cc}
         \partial_0+\partial_1 & -\partial_2-i\partial_3 \\
         -\partial_2+i\partial_3 & \partial_0-\partial_1 \\
       \end{array}
     \right)$$
Clearly $ \frac12\{{\mathcal{D},\mathcal{D}^*\}_M}={\mathcal{D}}{\mathcal{D}}^{*}=\partial^{2}_{0}-{\nabla}^2$.
Moreover, if we define $\chi^{\prime}=S(\Lambda)\chi$ and $\eta^{\prime}=T(\Lambda)\eta =(S^{\dag})^{-1}(\Lambda)\eta$ then equations (\ref{MJ}) are covariant under $SL(2,\mathcal{C})$ which means 
\begin{equation} \sigma^{\mu}\partial^{\prime}_{\mu}\eta^{\prime}=-m\chi^{\prime}\qquad {\rm and}\qquad
\sigma^*_{\mu}\partial^{\prime \mu}\chi^{\prime}=-m\eta^{\prime} \end{equation}
This is equivalent to the Lorentz invariance of the Klein-Gordan equation $$(\partial^{2}_{0}-{\nabla}^2)\psi=m^2\psi$$ which can be factored into (\ref{MJ}).  \\

\noindent Note also that $[\mathcal{D},{\mathcal{D}^*}] = $ 0  . We refer to the pair $(X, X^*)$ as Majorana fields. 

\subsection{Pauli Products} 
Noting that $$X^{\prime}X^{\prime*}=XX^{*}=x^2_0-x^2_1-x^2_2-x^2_3$$ 
and that for any (pseudo) inner product $\left<X+Y,X+Y\right>=(X+Y)(X^*+Y^*)$ is a scalar, it follows from the bi-linearity and symmetry that
$$\left<X+Y,X+Y\right>=\left<X,X\right> + 2\left<X,Y\right>+\left<Y,Y\right>$$ and therefore 
$$2\left<X,Y\right> = XY^*+YX^*$$
Equivalently, we can define a Pauli inner product  by\footnote{Apart from a notation change, this expression is identical with equation (20) of Pauli's original paper \cite{pauli}, applied to what Pauli refers to as the Jordan and Wigner bracket.}      
\begin{eqnarray} \left<X,Y\right>_P &\equiv&\frac14(\{X,Y^*\}+\{Y,X^*\})\\ 
&=&\frac14(XY^*+Y^*X+YX^*+X^*Y )\\
&=&x_0y_0-x_1y_1-x_2y_2-x_3y_3 =\left<x,y\right>\label{FD}\end{eqnarray}
which is clearly Lorentz invariant. Indeed, it is invariant under $SL(2,\cal{C})$ (and not just $SU(2,\cal{C}))$, provided we agree that conjugacy is preserved according to the rule established in the previous lemma. In other words if $T=(S^{\dag})^{-1}$,   
\begin{eqnarray*} \{X^{\prime},Y^{\prime}\}_P&\equiv& \frac14(SXS^{\dagger}TY^*T^{\dag}+TY^*T^{\dagger}SXS^{\dag})\\
&& \qquad +\frac14(SYS^{\dagger}TX^*T^{\dag}+TX^*T^{\dagger}SYS^{\dag})\\
&=& \frac14(SXY^*T^{\dag}+TY^*XS^{\dag})+\frac14(TX^*YS^{\dag}+SYX^*T^{\dag}) \\
&=&\frac14(S(XY^*+YX^*)T^{\dag}+T(X^*Y+Y^*X)S^{\dag})\\
&=&\frac12 S(x_0y_0-x_1y_1-x_2y_2-x_3y_3)T^{\dag}\\
&&\qquad +\frac12 T(x_0y_0-x_1y_1-x_2y_2-x_3y_3)S^{\dag}\\
&=&\{X,Y\}_P
\end{eqnarray*}
Therefore the Pauli inner product preserves conjugacy. 

By making a slight modification\footnote{Pauli's equation for commutators is symmetric and contains the expression $[X,Y^*]+[Y,X^*]$. He failed to note that it is always 0 for all $X$ and $Y$ regardless of the statistics. In contrast, we write $[X,Y^*]+[X^*,Y]$ which is anti-symmetric. } of equation (20) of article (\cite{pauli}), we can also define a Pauli outer product  by
\begin{equation} [X,Y]_P\equiv \frac12([X,Y^*]+[X^*,Y])=\frac12(XY^*-Y^*X+X^*Y-YX^* )\end{equation}
This reduces with a little algebra to 
\begin{equation} [X,Y]_P= -[\tilde{X},\tilde{Y}]\end{equation}
where $\tilde{X}=x_1\sigma_1+x_2\sigma_2+x_3\sigma_3$ is a vector in three dimensional space. Note that this also means $[X,X^*]=[X,X]=0.$

In particular,  in the case of a  singlet state, $\tilde{X}$ and $\tilde{Y}$,  $[\tilde{X},\tilde{Y}]\ne 0$. It also follows from equations (9) - (11) that in the case of singlet states $$ \{\tilde{X},\tilde{Y}\}_P=0 \ ,$$ while in the case of non-singlet states $$\{\tilde{X},\tilde{Y}\}_P<0 \ .$$ In other words, two particles cannot simultaneously obey Fermi-Dirac and Bose-Einstein statistics.  

\subsection{Restricted tensor products}
In Theorem 1, we let $V=V_1\otimes \dots \otimes V_n$, where each $V_i$ is an n-dimensional vector space, which means we chose the dimension of each vector to be the same value as the number of tensor products of the space itself. Specifically in the case of a two dimensional Euclidean space, the wedge product of two vectors ${\bf x}=(x_1,x_2)$ and $ {\bf y}=(y_1,y_2)$
generates a singlet state:
$$2{\bf x}\wedge{\bf y}={\bf x}\otimes{\bf y}- {\bf y}\otimes{\bf x}=(x_1y_2-y_1x_2){\bf e}_1\wedge{\bf e}_2$$
where ${\bf e}_1$ and ${\bf e}_2$ represent respectively the unit vectors $(1,0)$ and $(0,1)$. On the other hand, if we switch to a spinor formulation and use the Pauli spin matrices as a basis then we can identify
$${\bf x}\leftrightarrow X=\left(
       \begin{array}{cc}
       x_1 & x_2 \\
        x_2 & -x_1 \\
       \end{array}
     \right)\qquad {\rm and}\qquad {\bf y}\leftrightarrow Y=\left(
       \begin{array}{cc}
       y_1 & y_2 \\
        y_2 & -y_1 \\
       \end{array}
     \right)$$
A direct calculation gives $[X,Y]=2(x_1y_2-y_1x_2)\sigma_1\sigma_2$. It is worth noting that in the case of two dimensional Majorana fields $[X,X^*]=0$. In other words, Majorana fields can never form a singlet state and consequently never obey (by Theorem 1) Femi-Dirac statistics.

To complete the theory, we need to extend our results to Minkowski space. With this in mind, let $x,\ y \in {\mathcal R}^1_3$ and consider the tensor product $x\otimes y=x^iy^je_i\otimes e_j$ and a linear map $\phi( x\otimes y)=x^iy^j\sigma_i\sigma_j$. This is equaivalent to identifying for $i\ne j$
$$e_0\otimes e_j=e_j\otimes e_0\ {\textrm and}\ e_i\otimes e_j=-e_j\otimes e_i$$
In keeping with the Pauli outer product defined above, we define the conjugate wedge product by $${\bf x}\wedge_c {\bf y}\equiv \frac12({\bf x}\wedge{\bf  y}^* + {\bf x^*}\wedge {\bf y})\ .$$  Now let ${\bf x}=x^je_k\equiv x_0+\tilde{x}$, where $\tilde{x}=x_1e_1+x_2e_2+x_3e_3$ then this reduces to the bi-vector

\begin{equation} {\bf x}\wedge_C {\bf y}= - \tilde{x}\wedge \tilde{y}\end{equation}\newline
Clearly we can identify $4 ({\bf x}\wedge_C {\bf y}) \leftrightarrow [X,Y]_P$

Moreover, from basic geometry, we can see that any linear transformation with an eigenvector $\tilde{z}$ that is orthogonal to the plane spanned by  $\tilde{x}$ and $\tilde{y}$  will be such that $  \tilde{x}\wedge \tilde{y}$ remains invariant in accordance  with  Corollary 1. On the other hand those rotations that shift the eigenvector will not remain invariant.  However, the norm of the bivector  will remain invariant under $SL(2,\cal{C})$.

\section {Conclusion}
Based on the above, it should be clear that a necessary and  sufficient condition for the Pauli exclusion principle to be valid 
is the requirement that the quantum state of a system of $n$ particles be invariant under the action of the  $SL(n,{\cal C}$) group. As we have already noted this requires the existence of spin singlet states, which means that spin entanglement is a necessary requirement to exhibit Fermi-Dirac statistics.  Indeed,  Fermi-Dirac statistics can be defined as the statistics of n indistinguishable singlet states \cite{ohara}.

\end{document}